\documentclass[twocolumn,prb,superscriptaddress,10pt]{revtex4-1}

\usepackage{amsfonts}
\usepackage{amsmath}
\usepackage{amssymb}
\usepackage{graphicx}
\usepackage{hyperref}
\usepackage{latexsym}
\usepackage{color}

\begin{document}

\title{Cyclotron orbit knot and tunable-field quantum Hall effect}

\author{Yi Zhang}
\email{frankzhangyi@gmail.com}
\affiliation{International Center for Quantum Materials, Peking University, Beijing, 100871, China}
\affiliation{School of Physics, Peking University, Beijing, 100871, China}
\affiliation{Department of Physics, Cornell University, Ithaca, NY 14853, USA}
\affiliation{Kavli Institute for Theoretical Physics, University of California, Santa Barbara, CA 93106, USA}

\begin{abstract}
From a semi-classical perspective, the Bohr-Sommerfeld quantization of the closed cyclotron orbits for charged particles such as electrons in an external magnetic field gives rise to discrete Landau levels and a series of fascinating quantum Hall phenomena. Here, we consider topologically nontrivial physics from a distinct origin, where the cyclotron orbits take nontrivial knotting structures such as a trefoil knot. We present a scenario of a Weyl semimetal with a slab geometry, where the Fermi arcs on the opposing surfaces can cross without interfering with each other and form a knot together with the bulk Weyl nodes, and in an external magnetic field, the resulting chiral Landau levels. We provide a microscopic lattice model to realize a cyclotron orbit with an unconventional geometry of a trefoil knot and study the corresponding quantum oscillations. Interestingly, unlike the conventional ring-shaped cyclotron orbit, a trefoil knot is self-threading, allowing the magnetic field line along the novel cyclotron orbit to contribute to the overall Berry phase, therefore altering the external magnetic field for each quantization level. The cyclotron orbit knot offers a new arena of the nontrivial knot theory in three spatial dimensions and its subsequent physical consequences.
\end{abstract}

\maketitle

Integer quantum Hall effect\cite{IQHE1980} and the subsequently-discovered zoo of miscellaneous topological phases\cite{TsuiFQH, LaughlinFQH, Haldane1988, Kane2005, FuKaneMele3Dti, TI2010, Chen2012, Wan2011, Xu2015} represent an active research area in condensed matter physics. The physics idea, however, may trace back further to the insightful semiclassical quantization of the cyclotron orbits\cite{Onsager1952, Lifshitz1956}. Following the semiclassical equations of motion, charged particles such as electrons cycle in the plane normal to the magnetic field around the constant energy contours, which then become quantized according to the Bohr-Sommerfeld quantization condition. In two dimensions, all electron degrees of freedom are quenched into these cyclotron orbits with discrete and degenerate energy values - the Landau levels - in a magnetic field. The spacing and degeneracy of the Landau levels are proportional to the magnetic field strength, giving rise to the oscillations in material electronic properties as a function of the applied magnetic field such as De Haas–van Alphen effect, Shubnikov–de Haas effect and other fascinating facets of the quantum Hall effect.

\begin{figure}
\includegraphics[scale=0.6]{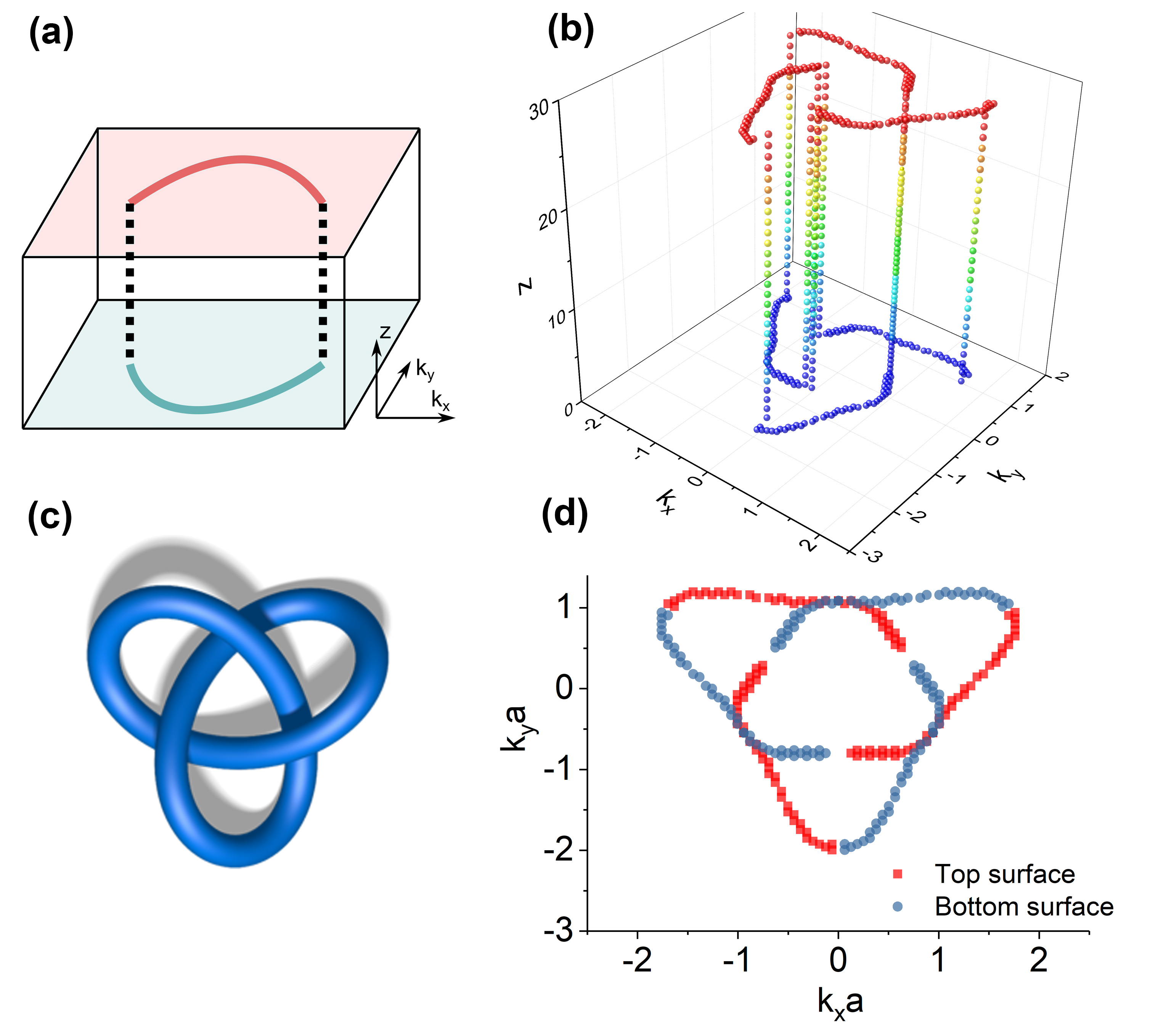}
\caption{(a) The Weyl orbit consists of the Fermi arcs on the top (red) and bottom (blue) surfaces and the chiral Landau levels from the Weyl nodes in the bulk (black dashed lines). This Weyl orbit is still topologically equivalent to a conventional ring-shaped cyclotron orbit. (b) The Fermi arcs of the Weyl semimetal model in Eq. \ref{eq:ham} on the top and bottom surfaces of a slab and the bulk tunneling at the six Weyl nodes assemble a cyclotron orbit with a trefoil-knot geometry. For a better visibility, the points show the values of $\left(k_x, k_y, z\right)$ for low-energy states within an energy window $E\in \left[-0.1,0.1\right]$ around the Weyl nodes. In the presence of a magnetic field along $\hat z$, the shape of the cyclotron orbit in three spatial dimensions is related by a 90-degree rotation in the $xy$ plane. (c) A trefoil knot is a closed loop with three crossings and topologically distinctive from a ring-shaped loop. (d) The projection of panel (b) onto the $k_x$, $k_y$ plane demonstrates the three crossings similar to those in panel (c). Thanks to the spatial separation between the surfaces, the Fermi arcs manage to cross without interfering with each other.}
\label{fig:trefoil}
\end{figure}

A relatively young sibling in the topological material family is the topological Weyl and Dirac semimetals in three dimensions\cite{Wan2011, Xu2015}. Around the Weyl nodes - selected points in their Brillouin zone, the low-energy electronic excitations disperse linearly, resembling the Weyl fermions in models for high energy physics. Also, the surface electronic states consist of exotic open Fermi arcs\cite{Wan2011, Xu2015}. In the presence of a magnetic field, the Weyl fermions become quantized as the chiral Landau levels that disperse along or against the magnetic field depending on the Weyl fermions' chirality, exhibiting the chiral anomaly phenomenon\cite{Nielsen1983}. These chiral Landau levels, together with the Fermi arcs on the top and bottom surface in a slab geometry, assemble a novel type of cyclotron orbits in the Dirac and Weyl semimetals, dubbed as the Weyl orbit\cite{Potter2014, frank2016}, see Fig. \ref{fig:trefoil}(a) for illustration. The Weyl orbit extends in and promotes the Landau quantization to three spatial dimensions, and the corresponding quantum oscillations\cite{Moll2016} and quantum Hall transports\cite{Wang2017, Zhang2017, Zhang2018} have been established experimentally.

The search for topological phenomena in condensed matter physics\cite{Witten1989, Chen2012} has received much inspiration from the mathematical studies on topology such as knot theory, which investigates the nonequivalent classes of closed loops and the corresponding invariants, e.g., the Jones polynomial, in higher dimensional spaces. Indeed, a series of two-dimensional topological orders are characterized by topological quantum field theory on nontrivial quasi-particle world-line knots in 2+1-dimensional space-time\cite{Witten1989}. Also, the topologically distinct fermionic excitations in the form of connected links and knots in momentum space are discovered in nodal link\cite{Nodallink, Haizhou2017, Wangzhong2017, Ezawa2017, Zahid2017} and nodal knot\cite{Ezawa2017, Nodalknot} metals. In this work, we present our discovery of a new topology origin of quantum materials and phenomena, where the cyclotron orbit employs a nontrivial knot topology, such as a trefoil knot (Fig. \ref{fig:trefoil}(c)), in three spatial dimensions. Commonly, a cyclotron orbit is self-evading, since the couplings between nearby Fermi surfaces generally widen the gap and move them further apart, especially in a magnetic field, making crossings - a key ingredient of knots - difficult to realize. The Weyl orbits in the Dirac and Weyl semimetal slabs offer a solution to this difficulty, as the Fermi arcs on the top and bottom surfaces are spatially separated and can form crossings without interfering with each other. We provide a microscopic lattice model example where we realize a Weyl orbit with a trefoil knot geometry, see Fig. \ref{fig:trefoil}(b) and (d)\footnote{In comparison, the nodal link and nodal knot metals rest their topological distinctions on the nodal line topology in three-dimensional momentum space. They remain gapless in the presence of a magnetic field and disperse in the momentum space along the magnetic field. Therefore, their nodal links and knots do not translate to their cyclotron orbits in three spatial dimensions.}. In three spatial dimensions, such a cyclotron orbit knot is not adiabatically connected to a conventional ring-shaped cyclotron orbit including the conventional Weyl orbit in Fig. \ref{fig:trefoil}(a).

The nontrivial topology of the cyclotron orbit knot also has profound physical consequences. For instance, unlike a ring-shaped loop, a knot is self-threading. Therefore a magnetic field line along a cyclotron orbit knot contributes nontrivially to the overall Berry phase around the orbit. Tuning this flux allows the contribution from other sources such as the magnetic field to differ in order to reach a specific Landau quantization. Using our microscopic lattice model, we study the behavior of the quantum oscillations associated with the cyclotron orbit knot. In particular, we introduce a perturbation that creates an effective magnetic field that aligns with the electronic velocity and study its impact on the subsequent quantum Hall effect.

Without loss of generality, we consider an electronic tight-binding model on a three-dimensional hexagonal lattice for concreteness:
\begin{eqnarray}
H &=& \sum_{\left\langle ij\right\rangle,z , s} t \left(-1\right)^z c^\dagger_{jzs} c_{izs}
+ \sum_{\left\langle\left\langle ik\right\rangle\right\rangle, z, s} t'_{ik}\left(-1\right)^z  \sigma^z_{ss} c^\dagger_{kzs}  c_{izs}  \nonumber \\ &+&\sum_{i, \left\langle zz'\right\rangle,s} t_z \sigma^z_{ss} c^\dagger_{iz's} c_{izs}
+\sum_{\left\langle ik \right\rangle,\left\langle zz'\right\rangle, s} t''_{izkz'} \left(-1\right)^z c^\dagger_{kz's}  c_{izs} \nonumber \\
&+&\sum_{i,z'=z+1,s,s'} i\Delta \sigma^x_{ss'} \left(c^\dagger_{iz's'}c_{izs}-c^\dagger_{izs}c_{iz's'}\right)   \nonumber\\
&-&\sum_{i,z,s} \mu_{s}\left(-1\right)^z c^\dagger_{izs} c_{izs}
\label{eq:ham}
\end{eqnarray}
where $i,j,k$ are the coordinates in the $xy$ plane and $\bf \sigma$ are the Pauli matrices defined on the two pseudo-spins $s,s'=\uparrow, \downarrow$. The first two terms are hoppings in the $xy$ plane between the nearest neighbors and the next nearest neighbors, $t'_{ik}=\pm it'$, and the next two terms are near-neighbor hoppings between the nearest layers, $t''_{izkz'}=\pm it'_z$, where the $\pm i=\exp\left(3i\phi_{ik}\right)$ phases depend on the azimuthal angle $\phi_{ik}$ from $i$ to $k$\footnote{See the illustrative plot and model construction guidelines in Supplemental Materials.}. The last two terms are a coupling between the two pseudo-spins $s,s'=\uparrow,\downarrow$ and a chemical potential. In the rest of the paper, we set $t=-1.0$, $t'=-0.5$, $t_z=-1.0$, $t'_z=-0.5$, $\Delta=0.4$, $\mu_\uparrow=-5.1$, and $\mu_\downarrow=-3.5$ unless noted otherwise. All lattice constants are set to 1.

First of all, this model system is a Weyl semimetal. To see this, we transform Eq. \ref{eq:ham} into the momentum space $H=\sum_{\bf k} H_{\bf k}$, with
\begin{eqnarray}
H_{\bf k} &=& \tau^z \left[\epsilon_0(k_\perp) + \sigma^z\epsilon_z(k_\perp)\right] + \tau^x \sigma^z 2t_z\cos k_z  \\\nonumber
&+&\tau^y 4t'_z \sin k_z \left(\sin k_1+\sin k_2 + \sin k_3\right) - \tau^x\sigma^x 2\Delta \sin k_z
\label{eq:hamk}
\end{eqnarray}
where $\tau$ are the Pauli matrices on the even-odd layers. $k_z\in\left[0,\pi\right)$ is the momentum along the $\hat z$ direction, and
$k_1=k_x$, $k_2=\left(-k_x+\sqrt{3}k_y\right)/2$, $k_3=\left(-k_x-\sqrt{3}k_y\right)/2$ are momenta in the $xy$ plane. $\epsilon_0(k_\perp) + \sigma^z\epsilon_z(k_\perp)$ are the Fourier transform of the $t$, $t'$ and $\mu_s$ terms.
There are six Weyl nodes on the $k_z=\pi/2$ plane at $(k_x,k_y)=(0, -1.96)$ with $\sigma_z\simeq \uparrow$ and $(k_x,k_y)=(0, -0.834)$ with $\sigma_z\simeq \downarrow$ and their counterparts after the $C_3$ rotations\footnote{The $C_3$ rotation symmetry is for simplicity and not a necessary ingredient of the cyclotron orbit knotting geometry.}. The band gap closes at these Weyl nodes, and the low-energy dispersion around them is linear. Interestingly, for a model system with a finite thickness $L_z$ along the $\hat z$ direction, the surface Fermi arcs consist of three Fermi arcs on the top surface and three on the bottom surface. The $\left(k_x, k_y, z\right)$ location of the Fermi arcs and the bulk states at the Weyl nodes are shown in Fig. \ref{fig:trefoil}(b) for $L_z=29$, thick enough to separate the Fermi arc states on the opposing surfaces. Therefore, the constant energy contour employs a trefoil knot [Fig. \ref{fig:trefoil}(c)], which engages all six Weyl nodes and six Fermi arcs at once.

In the presence of an applied magnetic field along $\hat z$, the electrons cycle around the cyclotron orbit, whose shape in three spatial dimensions is related by a 90-degree rotation in the $xy$ plane. Therefore, we have established a cyclotron orbit with a trefoil knot geometry that is topologically distinctive and not adiabatically connected with the conventional ring-shaped cyclotron orbits. We note the important role the Weyl orbit physics plays in attaining cyclotron orbit knots: the top and bottom Fermi arcs are spatially separated and can safely traverse each other without being gapped out even in the presence of a variable magnetic field. Together with the bulk chiral Landau levels that descend from the Weyl nodes and weave the two surfaces back together, we can form crossings that are the cornerstones for knots and links. The construction of trefoil-knot-shaped cyclotron orbit can be straightforwardly generalized to cyclotron orbits with more complex knots and links.

\begin{figure}
\includegraphics[scale=0.35]{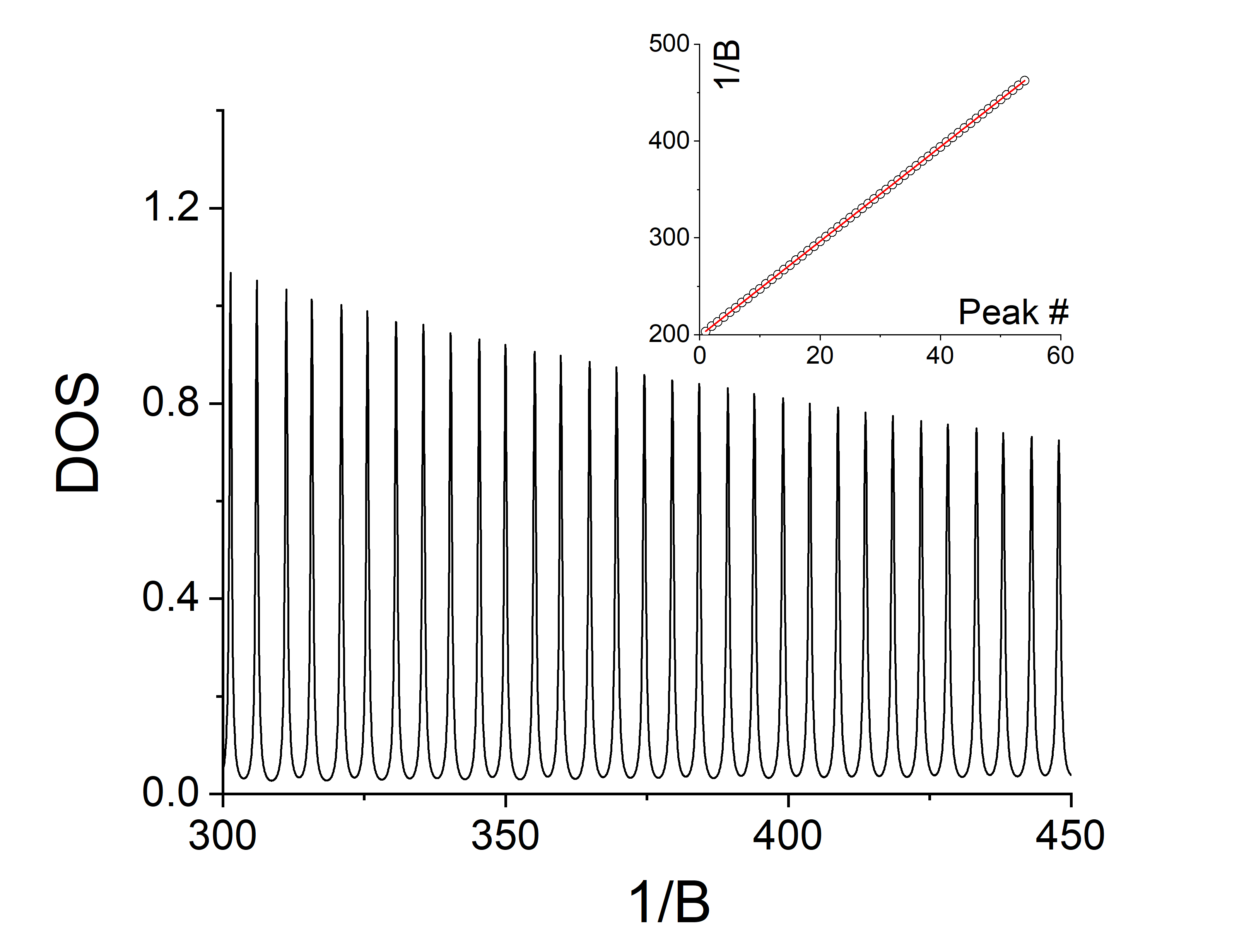}
\caption{The DOS versus the inverse magnetic field $1/B$ shows clear quantum oscillations with a single period. The Hamiltonian in Eq. \ref{eq:ham} with a magnetic vector potential $\vec{A}=(-By,0)$ is calculated using the recursive Green's function method. We focus on the chemical potential at the Weyl nodes and consider a slab with thickness $L_z=29$. A small imaginary part $\delta=0.001$ is added to the energy to account for a finite level width. Inset: a linear fit to the peak position within the $200<1/B<500$ range reveals a quantum oscillation period of $\Delta(1/B)\simeq 4.87$.}
\label{fig:qo1}
\end{figure}

\begin{figure}
\includegraphics[scale=1.5]{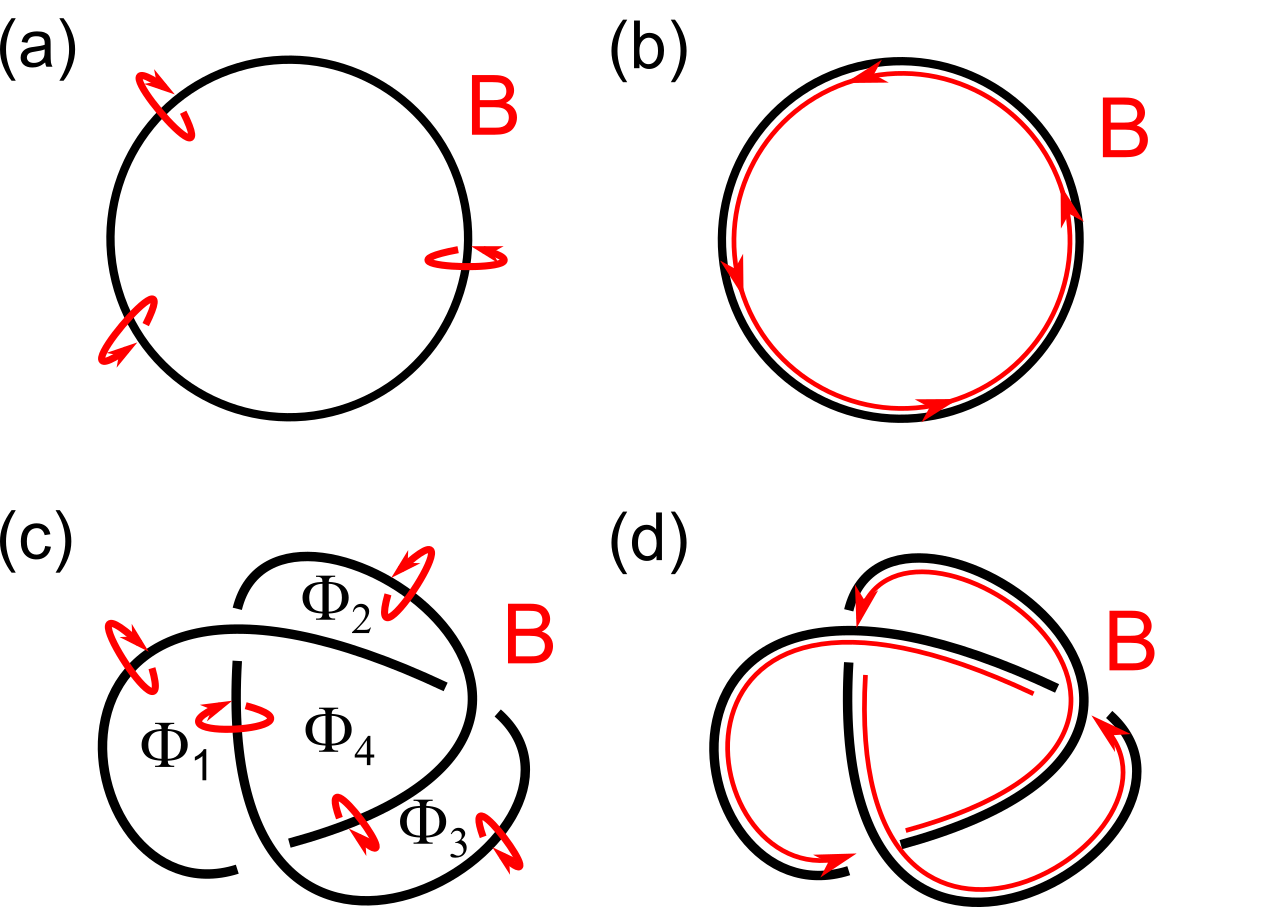}
\caption{The magnetic field line $\bf B$ (red arrows) for ring-shaped and trefoil-knot-shaped cyclotron orbits. (a) A magnetic field line through the ring contributes a Berry phase. (b) The magnetic field line along the ring-shaped cyclotron orbit has no Berry phase contribution. (c) The magnetic field line through the trefoil knot contributes a Berry phase. The contributions from the fluxes in different regions $\Phi_{total}=\sum_{i=1}^4 \alpha_i \Phi_i$ obey $\alpha_1=\alpha_2=\alpha_3=\alpha_4/2$. (d) Since a trefoil knot winds around itself, the magnetic field line along the cyclotron orbit knot now contributes a non-zero Berry phase.}
\label{fig:fluxknot}
\end{figure}

Next, we study the Landau quantization of the cyclotron orbit knot and consider the Hamiltonian in Eq. \ref{eq:ham} with a slab thickness of $L_z=29$ in the presence of a magnetic vector potential $\vec{A}=(-By, 0)$. Physical quantities such as the density of states (DOS) at the energy of the Weyl nodes can be obtained using the recursive Green's function method for sufficiently large system sizes. The results on the DOS versus the inverse magnetic field are summarized in Fig. \ref{fig:qo1}. We observe a single quantum oscillation period of $\Delta(1/B)=1/B_{n+1}-1/B_n\simeq 4.87$. Since the bulk chiral Landau levels are parallel to the magnetic field, the quantum oscillations of a Weyl orbit is determined by the area $S_k$ of the combined Fermi arcs from the top and bottom surfaces\cite{frank2016}. Interestingly, after projection and combination of the Fermi arcs [Fig. \ref{fig:trefoil}(d)], the area within the inner contour is $S_{k2}\sim 4.78\% \times S_{BZ}$ of the surface Brillouin zone area $S_{BZ}$, and the area within the outer contour (including $S_{k2}$) is $S_{k1}\sim 15.2\% \times S_{BZ}$. That $\Delta(1/B) \simeq \left(S_{k1}+S_{k2}\right)^{-1} = 5.0$ indicates the magnetic flux enclosed in the inner contour contributes to the overall Berry phase twice. Indeed, straightforward counting suggests that the cyclotron orbit encloses the inner region twice upon each cycle. More rigorously, the contributions to the overall Berry phase $\Phi_{total}=\sum_{i=1}^4 \alpha_i \Phi_i$ obeys $\alpha_1=\alpha_2=\alpha_3=\alpha_4/2$, since the adiabatic change as illustrated in Fig. \ref{fig:fluxknot}(c) of a magnetic field loop with flux $\Phi$ indicates the equivalence $\alpha_1 \Phi=\alpha_2\Phi=\alpha_3\Phi=\left(\alpha_4-\alpha_1\right)\Phi=\left(\alpha_4-\alpha_3\right)\Phi$. On that account, our numerical results on quantum oscillations are fully consistent with the trefoil knot geometry of the cyclotron orbit.

Then, we discuss an interesting physical consequence of the nontrivial knot geometry of a cyclotron orbit - the tunable-field quantum Hall effect. Conventionally, the magnetic field line along the ring-shaped cyclotron orbit does not contribute to the overall Berry phase [Fig. \ref{fig:fluxknot}(b)]. In contrast, a trefoil knot is self-threading, and thus the magnetic field line along the cyclotron orbit knot contributes nontrivially to the overall Berry phase, see Fig. \ref{fig:fluxknot}(d). By controlling the contribution from such flux, we can modify the Landau quantization condition for other sources of the Berry phase, such as the applied magnetic field. The conventional spin-orbit interaction introduces a $\vec v$-dependent effective magnetic field that couples to the electron spin; here, we need to introduce a $\vec v$-dependent effective magnetic field $\vec B_{eff}$ that couples to the electron orbital angular momentum. One of the electron semiclassical equations of motion $d\vec{r}/dt=\vec{v}\left(\vec{k}\right)$ indicates that the cyclotron orbit is parallel to electron velocity at every instance and guarantees an effective magnetic field $\vec B_{eff}\parallel \vec v$, long overlooked due to the presence of only $\vec{B}\times \vec{v}$ term in the equations of motion\cite{Onsager1952, Lifshitz1956, QianNiu2010}, is along the cyclotron orbit irrespective of the shape details and contributes to the tunable-field quantum Hall effect.

In the case of a knotting Weyl orbit, since the Fermi arcs locally responsible for the crossings are on the surfaces, we limit our attention to the effective magnetic fields in the $xy$ plane. For instance, along the $\hat y$ direction, we may introduce a $\vec v$-dependent effective magnetic field with a perturbation $H'\propto v_y\left(\vec k + \eta z \hat x\right) - v_y\left(\vec k - \eta z \hat x\right)\simeq 2i\eta z \left[x, v_y\right]$ for small $\eta$, where $v_y=i\left[y, H\right]$ is the electron velocity in the $\hat y$ direction. However, it leads to a $\sin\left(k_3\right)-\sin\left(k_2\right)$ azimuthal angle dependence and vanishes after taking into account the $-\frac{1}{2}\hat y\pm \frac{\sqrt{3}}{2}\hat x$ directions. Therefore, we introduce an additional factor of $\cos\left(k_1\right)$, which is $\sim 1$ and a good approximation near the Weyl nodes and the crossing point where $v_y$ is the most important. After similar treatment to the $-\frac{1}{2}\hat y\pm \frac{\sqrt{3}}{2}\hat x$ directions, we obtain in total:
\begin{eqnarray}
H'&=&  \sum_{\left\langle\left\langle ik\right\rangle\right\rangle,z'=z+1,s} t'_z \exp{\left(3i\phi_{ik}\right)} c^\dagger_{kz's} c_{izs}\times \eta z \left(-1\right)^z
+\mbox{h.c.}\nonumber\\
\label{eq:pertb}
\end{eqnarray}
For slowly varying $\eta z$, $H'=\sum_{\bf k} H'_{\bf k}$ takes a momentum-space form
\begin{eqnarray}
H'_{\bf k}&\simeq& \tau^y 4 t'_z \eta z  \\& &
\times \sin k_z \left[\sin (k_2-k_1)+\sin(k_3-k_2)+\sin(k_1-k_3)\right]\nonumber
\end{eqnarray}
that only has $\tau^y$ component and thus does not interfere with the quantum oscillation period, which is determined by the combined surface Fermi arcs\cite{frank2016} and the $\tau^z$ terms in the original $H_{\bf k}$ in Eq. \ref{eq:hamk}. However, $H'$ does have an interesting impact on the model physics.

\begin{figure}
\includegraphics[scale=0.35]{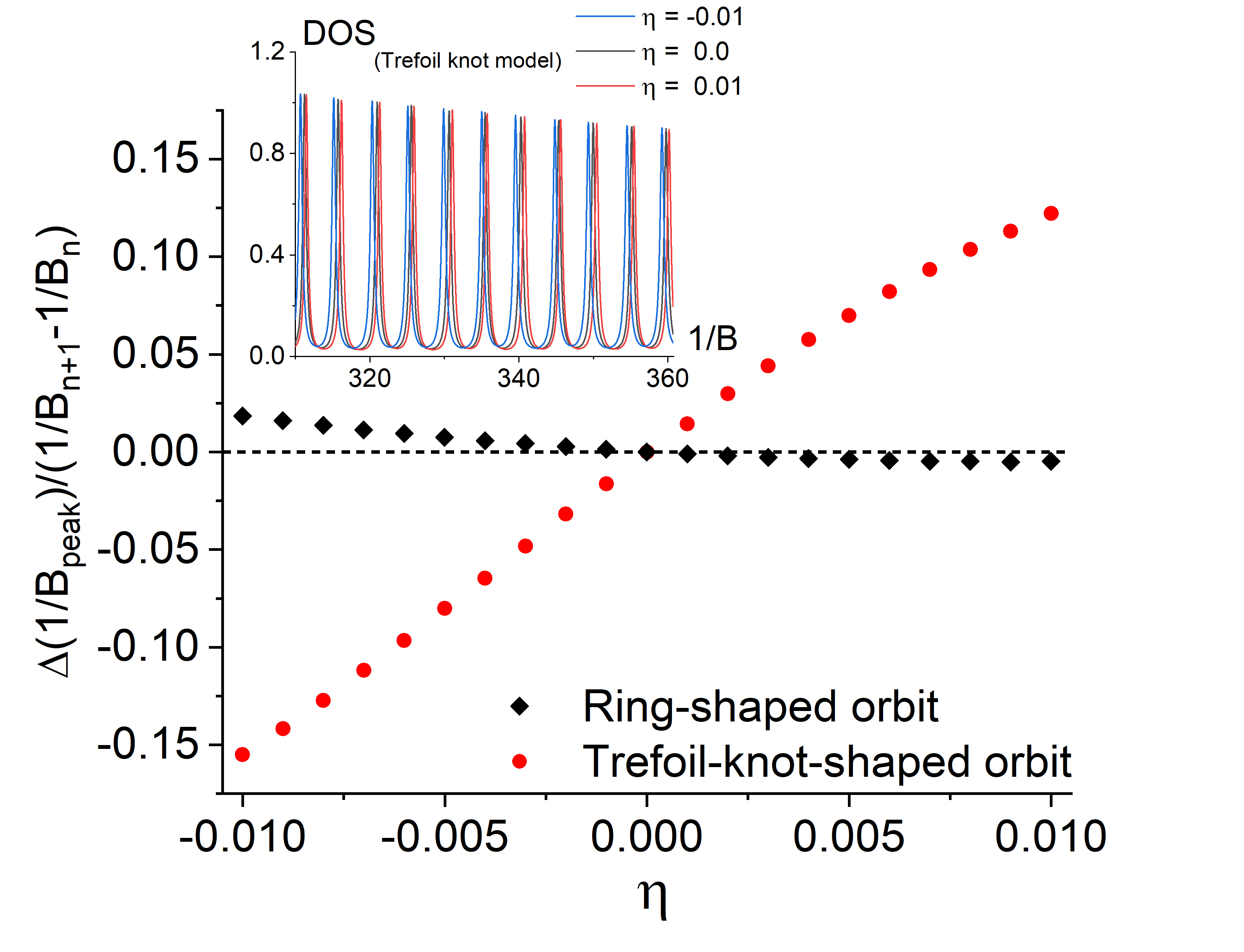}
\caption{The shifts of the DOS quantum oscillation peak positions $1/B_{\mbox{peak}}$ versus the amplitude $\eta$ of the applied perturbation in Eq. \ref{eq:pertb}. We track the DOS peak at $1/B\sim227.78$ at $\eta=0$ for the cyclotron orbit knot model and the DOS peak at $1/B\sim223.61$ at $\eta=0$ for the conventional Weyl orbit model as the $\sigma^z=\uparrow$ sector of $H+H'$ in Eq. \ref{eq:ham} and \ref{eq:pertb}. The perturbation creates an effective magnetic flux that contributes to the cyclotron orbit knots, and changes the Landau quantization condition for the magnetic field, see the red dots. In comparison, the $1/B$ quantization condition, shown as the black diamonds, hardly changes for the conventional ring-shaped Weyl orbit. Inset: the DOS peaks for the cyclotron orbit knot models displace with $\eta$ yet retain an identical quantum oscillation period of $\Delta\left(1/B\right)\simeq 4.87$ for $\eta=0, \pm0.01$. }
\label{fig:peakshift}
\end{figure}

As we include the perturbation $H'$ into the original Hamiltonian $H$ in Eq. \ref{eq:ham}, the magnetic field $B$ corresponding to each Landau level changes. We track the shift of the DOS quantum oscillation peak $\Delta\left(1/B_{\mbox{peak}}\right)=1/B_{\mbox{peak}}\left(\eta\right)-1/B_{\mbox{peak}}\left(\eta=0\right)$ as a function of $\eta$, and the results are summarized in Fig. \ref{fig:peakshift}. The ratio $\Delta\left(1/B_{\mbox{peak}}\right)/\left(1/B_{n+1}-1/B_n\right)$ measures the Berry phase contributed by the perturbation in unit of the magnetic flux quantum. On the other hand, the period of the quantum oscillations remains unchanged, see Fig. \ref{fig:peakshift} inset. Linear fits to 40 peak positions within the $200<1/B<400$ range indicate an identical period of $\Delta\left(1/B\right)\simeq 4.87$ for $\eta=0, \pm0.01, \pm0.02$. Therefore, the inclusion of $H'$ induces an \emph{extra} yet \emph{constant} phase to the cyclotron orbit knot.

In fact, the tunable-field quantum Hall effect is not fully unheard of - the overall Berry phase of the Weyl orbit receives a contribution from the bulk chiral Landau levels that depends on the Fermi energy, the tilting direction of the applied magnetic field, as well as the thickness of the system\cite{frank2016, Zhang2018}. The extra Berry phase in this work, however, is consequential to the nontrivial knotting topology of the cyclotron orbit and comes from a completely different origin: with or without the $H'$ perturbations, our model systems are still $C_3$ rotation symmetric, the Fermi energy is at the Weyl nodes, and the magnetic field is along $\hat z$ without tilting. Indeed, this contribution is unique to the cyclotron orbit knot model. In contrast, such an in-plane effective magnetic field incurs no orbital effect for conventional cyclotron orbits in two-dimensional electron systems. Further, we repeat the calculations for the $\sigma^z=\uparrow$ sector of the original Hamiltonian $H$ in Eq. \ref{eq:ham}, which describes a conventional Weyl semimetal with a topologically-ring-shaped Weyl orbit. Its Weyl orbit also consists of three pairs of chiral Landau levels and three pieces of Fermi arcs on each of the top and bottom surfaces. As we change the amplitude $\eta$ of the perturbation, given by the $\sigma^z=\uparrow$ sector of $H'$ in Eq. \ref{eq:pertb}, the DOS peaks in the quantum oscillations hardly shift, see the black diamond symbols in Fig. \ref{fig:peakshift}. The small deviations from ideal theory expectation may be due to the approximation in $H'$.

In conclusion, we have shown a new topological prospect where the cyclotron orbit takes a nontrivial knot geometry in three spatial dimensions and proposed the Weyl semimetal slab as an example of realization. We have also provided a microscopic lattice model, investigated its Landau quantization, and demonstrated that the unusual knotting topology of the cyclotron orbit allows the continuous tuning of the magnetic field condition for quantum Hall effect. We note that the Weyl orbits depend on the surface states, thus the cyclotron orbit knot can be realized on existing Weyl semimetals, such as the RhSi with large Fermi arcs\cite{largearcs}, via proper surface design\footnote{Please refer to Supplemental Materials for schematic potential realization.}.

Since the knots are topologically protected and invariant in three spatial dimensions, a cyclotron orbit knot is not adiabatically connected with the conventional ring-shaped cyclotron orbits\footnote{See further discussions on cyclotron orbit knot's stability and transition in Supplemental Materials.} and allowed to have characteristic topological properties. In topological quantum field theory, the nontrivial anyonic phases of quasi-particle braiding and statistics in two spatial dimensions are consequential to the nontrivial Jones polynomial of the knots between the quasi-particle world-lines in 2+1-dimensional space time\cite{Witten1989}. The consequences of cyclotron orbit knots in three spatial dimensions on characteristic properties, such as intrinsic Berry phase contributions\cite{QianNiu2010}, the scenarios beyond $U(1)$ gauge field, etc. and their connections to the knot invariants, are interesting further directions for future exploration.

Acknowledgment: YZ acknowledges support from the start-up grant at International Center for Quantum Materials, Peking University, the Bethe fellowship at Cornell University, and the National Science Foundation under Grant No. NSF PHY-1748958. The computation was supported by High-performance Computing Platform of Peking University.

\bibliography{refs}

\section{Cyclotron orbit knot model construction}

\begin{figure}
\includegraphics[scale=0.5]{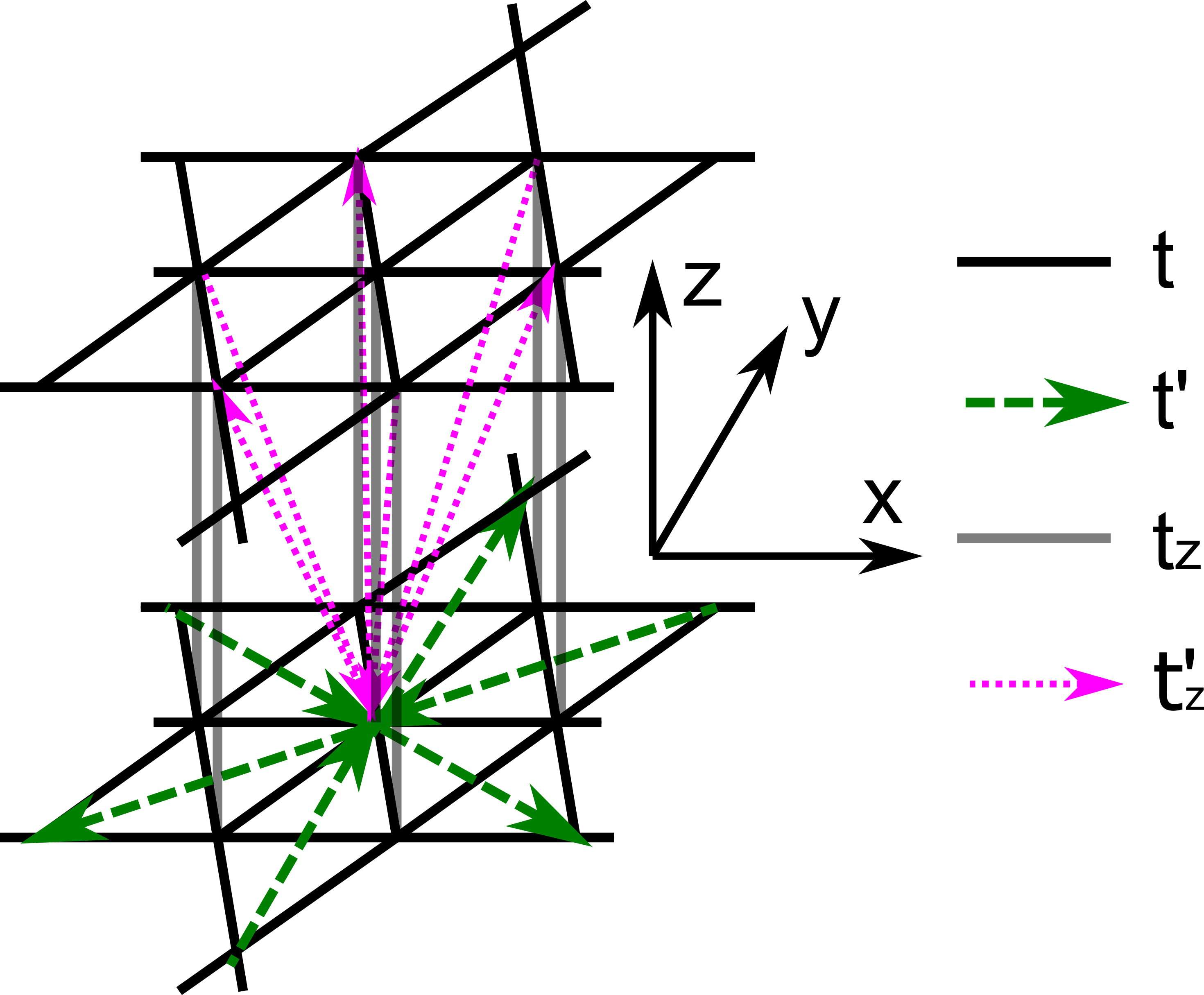}
\caption{Our Hamiltonian in Eq. (1) in the main text describes a tight-binding model on the hexagonal lattice with four types of near-neighbor hoppings. The arrows denote imaginary amplitudes where the phase is $i$ for hoppings along the arrow and $-i$ against the arrow.}
\label{fig:ham}
\end{figure}

\begin{figure}
\includegraphics[scale=1.2]{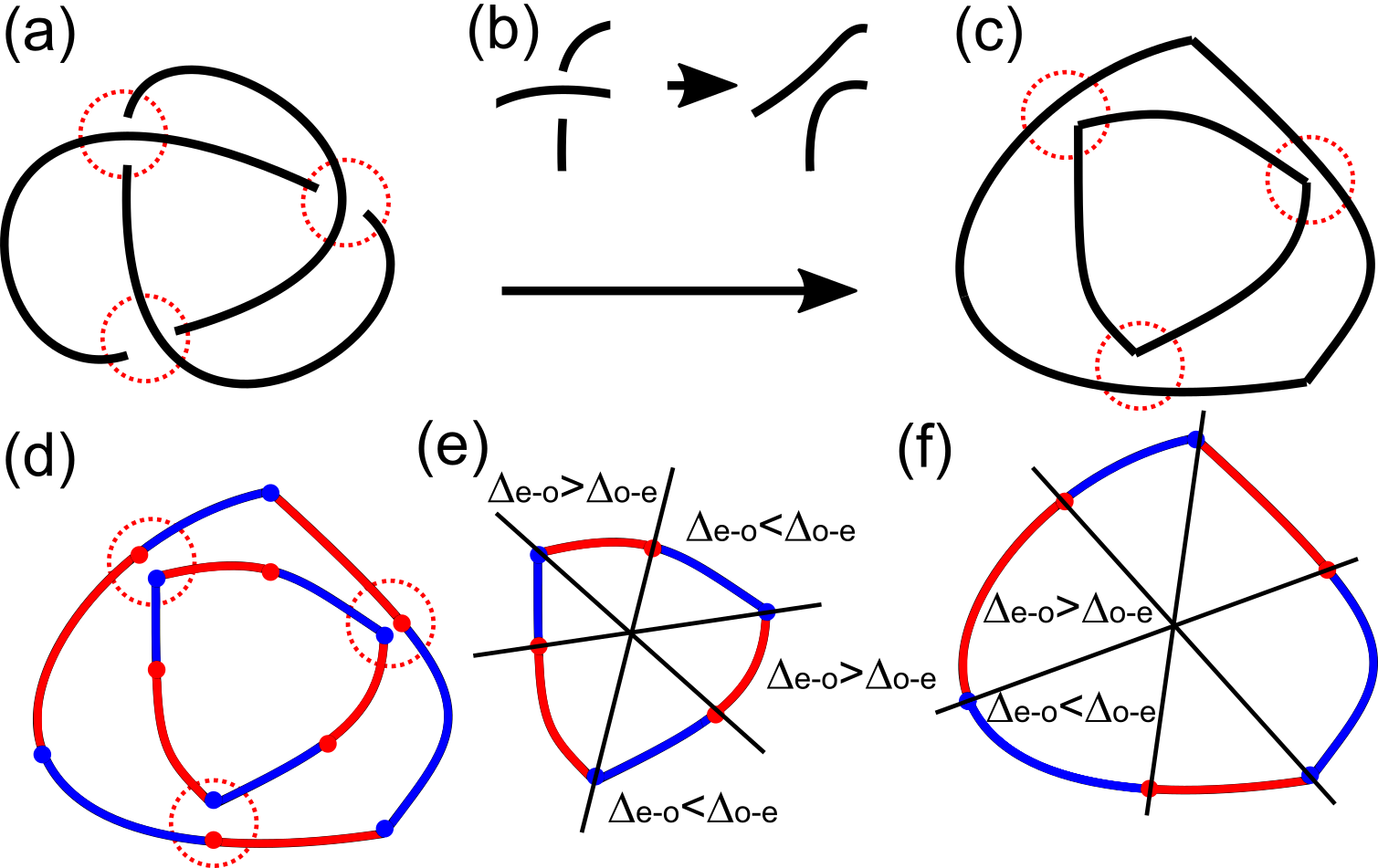}
\caption{The guideline for building a cyclotron orbit knot: (a) projecting the target knot or link onto a 2d plane results in a series of crossings. (b) By switching partners at each crossing, the diagram can be decomposed into several closed pockets as in (c), which can be realized as the constant energy contours of 2d systems. (d) Based upon a layered construction of the 2d systems, one can build a Weyl semimetal, where the red and blue curves are the Fermi arcs on the top and bottom surfaces projected onto the $\left(k_x, k_y\right)$ plane, respectively. The red and blue dots denote the projected locations of the Weyl nodes with opposite chirality. A coupling between the contours annihilates the circled pairs of nearby Weyl points and re-connects the crossings as in (a). (e) and (f) are an example of the inter-layer coupling in the layered construction that gives rise to the Fermi arcs in (d) for the inner contour and outer contour, respectively.}
\label{fig:app1}
\end{figure}

In the main text, we present a tight-binding model of a Weyl semimetal that hosts trefoil-knot-shaped Weyl orbit, see Fig. \ref{fig:ham} for an illustration of the model hoppings. In this section, we discuss the general guidelines for this and other cyclotron orbit models alike, via a layered construction as illustrated in Fig. \ref{fig:app1}. We assume the same slab geometry as in the main text and define the $\hat z$ direction to be the surface normal. We also use the $\left(k_x, k_y, z\right)$ coordinates, so that the knotting structure of the constant energy contour translates into that of the cyclotron orbit in three spatial dimensions in the presence of an external magnetic field in the $\hat z$ direction.

First, we project the target knot or link structure in its simplest form - a one-dimensional loop embedded in three dimensions - onto a two-dimensional plane. The projected diagram of a nontrivial loop necessarily involves a number of crossings, see Fig. \ref{fig:app1}(a). These crossings are the obstacles towards the conventional cyclotron orbit realization.

Next, we switch the partners at each crossing as in Fig. \ref{fig:app1}(b), so that the projected diagram becomes separate contours (Fig. \ref{fig:app1}(c)). These contours are realizable as the constant energy contours of 2d systems.

Then, we resolve the crossing issues via the Fermi arcs on the opposing surfaces of a Weyl semimetal slab. Fig. \ref{fig:app1}(d) is the projection of the Fermi arcs onto the $\left(k_x, k_y\right)$ plane, and the red and blue portions of the Fermi surfaces are localized on the top and the bottom surfaces, respectively. This top and bottom assignments should be consistent with that of the original crossings (Fig. \ref{fig:app1}(a)). Such a Weyl semimetal can be realized with a slab with $L_z=2L+1$ layers of the 2d system from the previous step and an inter-layer coupling that is stronger between the $2n$ layer and $2n+1$ layer in the regions containing the blue portions, and stronger between the $2n-1$ layer and $2n$ layer in the regions containing the red portions, respectively, see Fig. \ref{fig:app1}(e) and (f) for illustrations, $n\in \mathbb{Z}$. Weyl nodes emerge at the junctions between the red and blue portions.

Finally, we introduce coupling between the different contours, which gaps out the nearby Weyl nodes with opposite chirality, whose projected locations are denoted in Fig. \ref{fig:app1}(d) as blue and red dots, respectively. In particular, the pair of Weyl nodes circled out annihilates, and the connections at the crossings re-establish as the original contour of the targeted knot geometry.

It is also straightforward to see that for a slab with $L_z=2L$ layers, the Fermi surfaces on the top and bottom surfaces coincide, and the corresponding Weyl orbits are topologically equivalent to the conventional ring shape. Therefore, like the conventional Weyl orbits, the cyclotron orbit knot is sensitive to surface details and can be induced from the existing Weyl semimetal materials and models with proper surface design. On the other hand, the cyclotron orbit knot and the conventional cyclotron orbits are separated by Lifshitz transitions and their distinctions should remain relatively stable against small perturbations. We will further discuss these aspects in the next section.

\section{Cyclotron orbit knot transitions}

In this section, we study the stability of and transition between the cyclotron orbit knot and the conventional ring-shaped cyclotron orbits. We start with the model system in Eq. (1) in the main text on a slab of $L_z=2L$ layers and impose an additional potential $H_{top}=\epsilon \sum_{i,z=L_z,s} c_{izs}^\dagger c_{izs} $ on the topmost layer. As is in the main text, $i$ is the coordinate in the $xy$ plane, and $s$ is the pseudo-spin. Following the layer construction from the last section, the Weyl orbits at $\epsilon=0$ are conventional cyclotron orbits with similar Fermi arcs on the top and bottom surfaces. On the other hand, for large $\epsilon\gg 1$, the topmost layer is projected out leaving an odd number of layers, and the Fermi arcs on the top surface and the corresponding Weyl orbit resume the form of cyclotron orbit knots as displayed in the main text.

\begin{figure}
\includegraphics[scale=0.15]{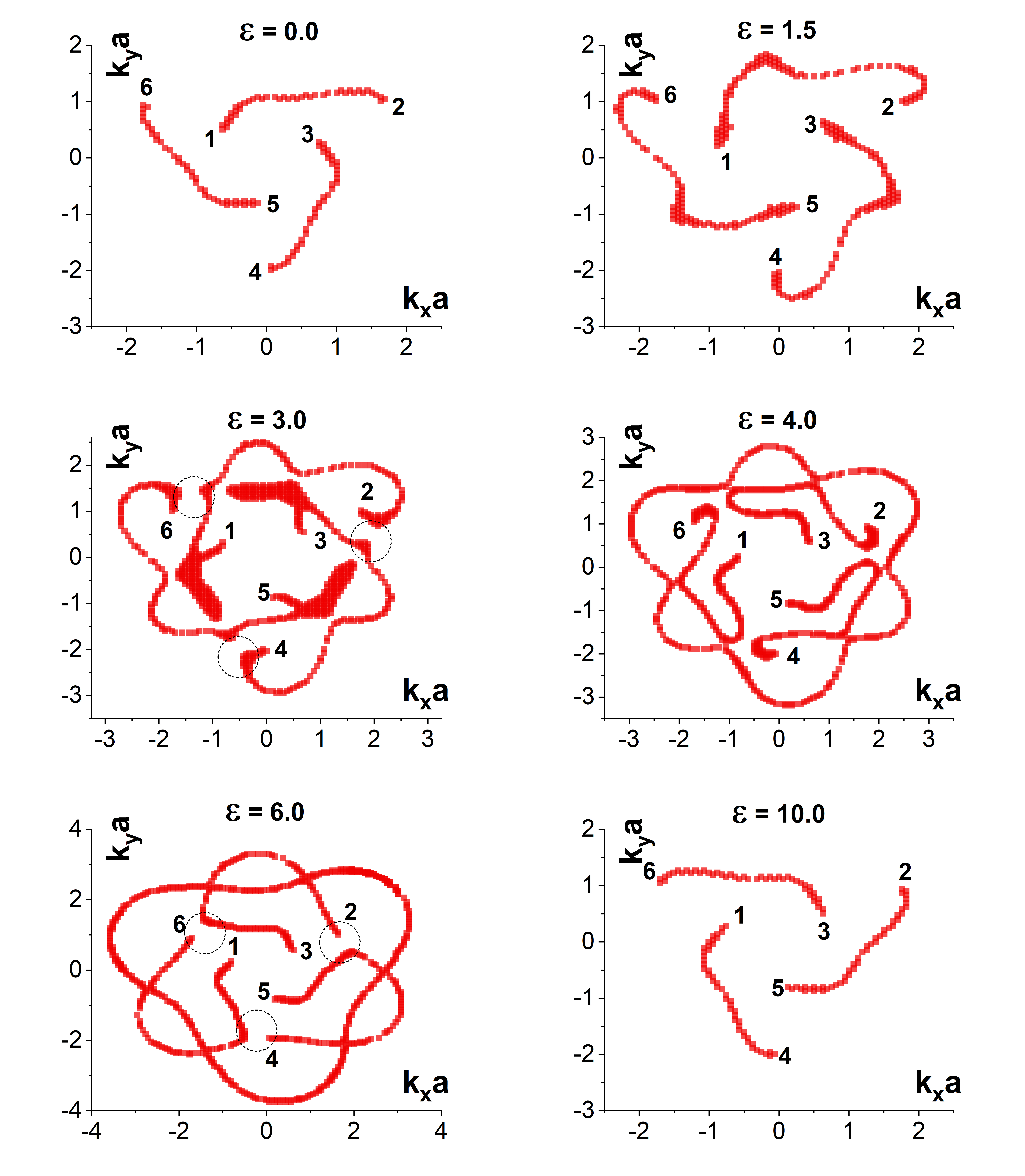}
\caption{The Fermi arcs on the top surface of a slab of $L_z=2L$ layers with an additional potential $H_{top}=\epsilon \sum_{i,z=L_z,s} c_{izs}^\dagger c_{izs}$ on the topmost layer. The Weyl nodes deep in the bulk and the Fermi arcs on the bottom surface remain unchanged during the process. The top surface Fermi arcs connectivity results in conventional cyclotron orbits at $\epsilon=0$ (an even number of layers) and cyclotron orbit knot at $\epsilon=10$ (effectively an odd number of layers). These two limits are separated by (a series of) Lifshitz transitions and there is a finite parameter range where the Fermi arcs connectivity and the cyclotron orbit topology remains intact. The black dashed circles denote the locations of the Lifshitz transitions at intermediate $\epsilon$ values.}
\label{fig:app2}
\end{figure}

\begin{figure}
\includegraphics[scale=0.275]{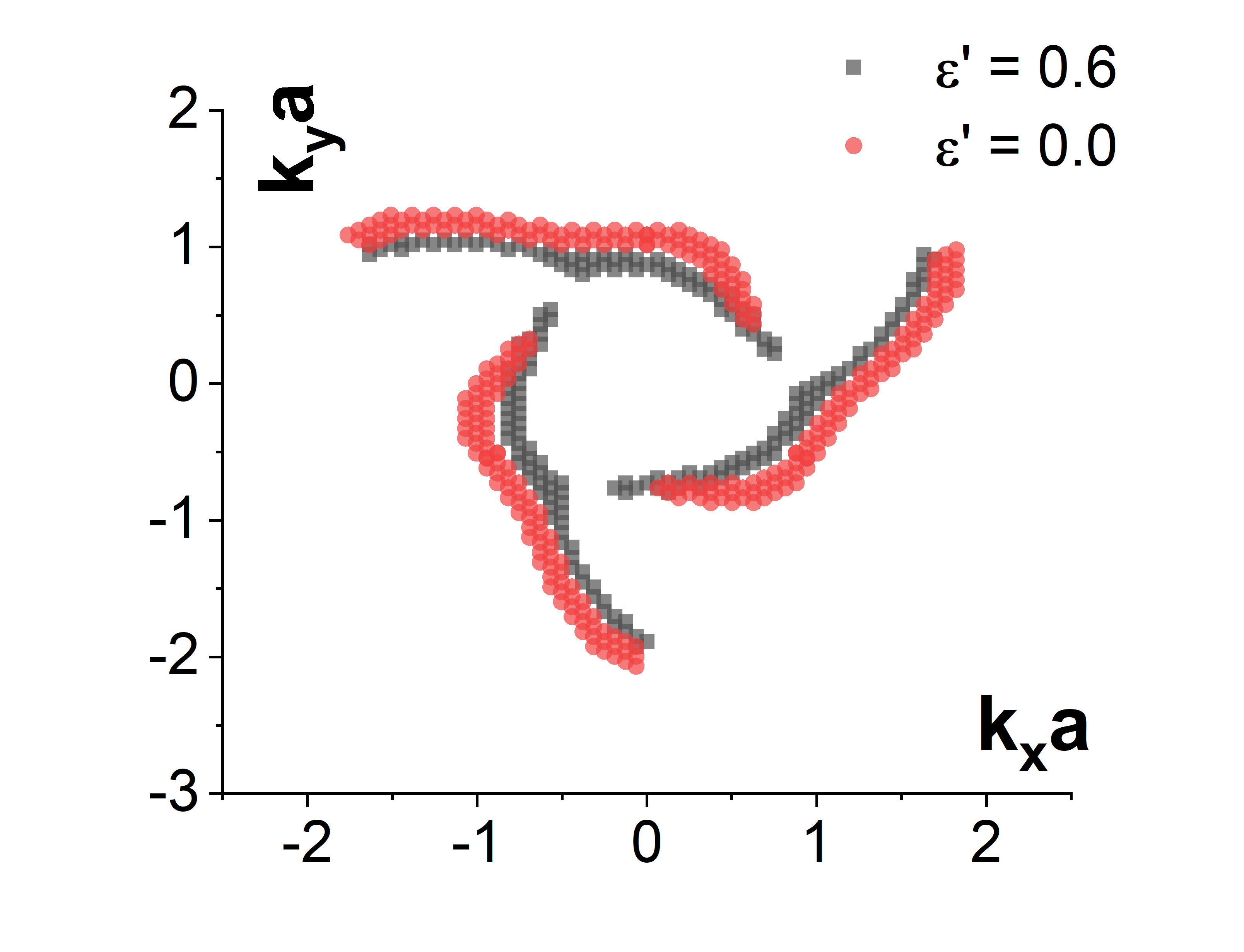}
\caption{The Fermi arcs on the top surface of a slab of $L_z=2L+1$ layers with an additional potential $H_{top}=\epsilon' \sum_{i,z=L_z,s} c_{izs}^\dagger c_{izs} $ on the topmost layer indicate that the corresponding cyclotron orbit knot is stable against small surface perturbations.}
\label{fig:app3}
\end{figure}

The results for various values of $\epsilon$ are summarized in Fig. \ref{fig:app2}. This is an example that by proper alteration of the surface property, we successfully change the top surface Fermi arcs connectivity and transform the conventional cyclotron orbits ($\epsilon=0$) into nontrivial cyclotron orbit knots ($\epsilon=10$). In addition, we see that both types of cyclotron orbits are stable within a finite range of parameters and separated by Lifshitz transitions at intermediate values of $\epsilon$.

As another example, we start with the model system in Eq. (1) in the main text on another slab of $L_z=2L+1$ layers and impose an additional potential $H_{top}=\epsilon' \sum_{i,z=L_z,s} c_{izs}^\dagger c_{izs} $ on the topmost layer. The resulting cyclotron orbit knot is clearly stable against finite perturbations, see Fig. \ref{fig:app3}. At the same time, the bulk is a topological Weyl semimetal whose chiral Landau levels are also protected against perturbations.

\begin{figure}
\includegraphics[scale=0.75]{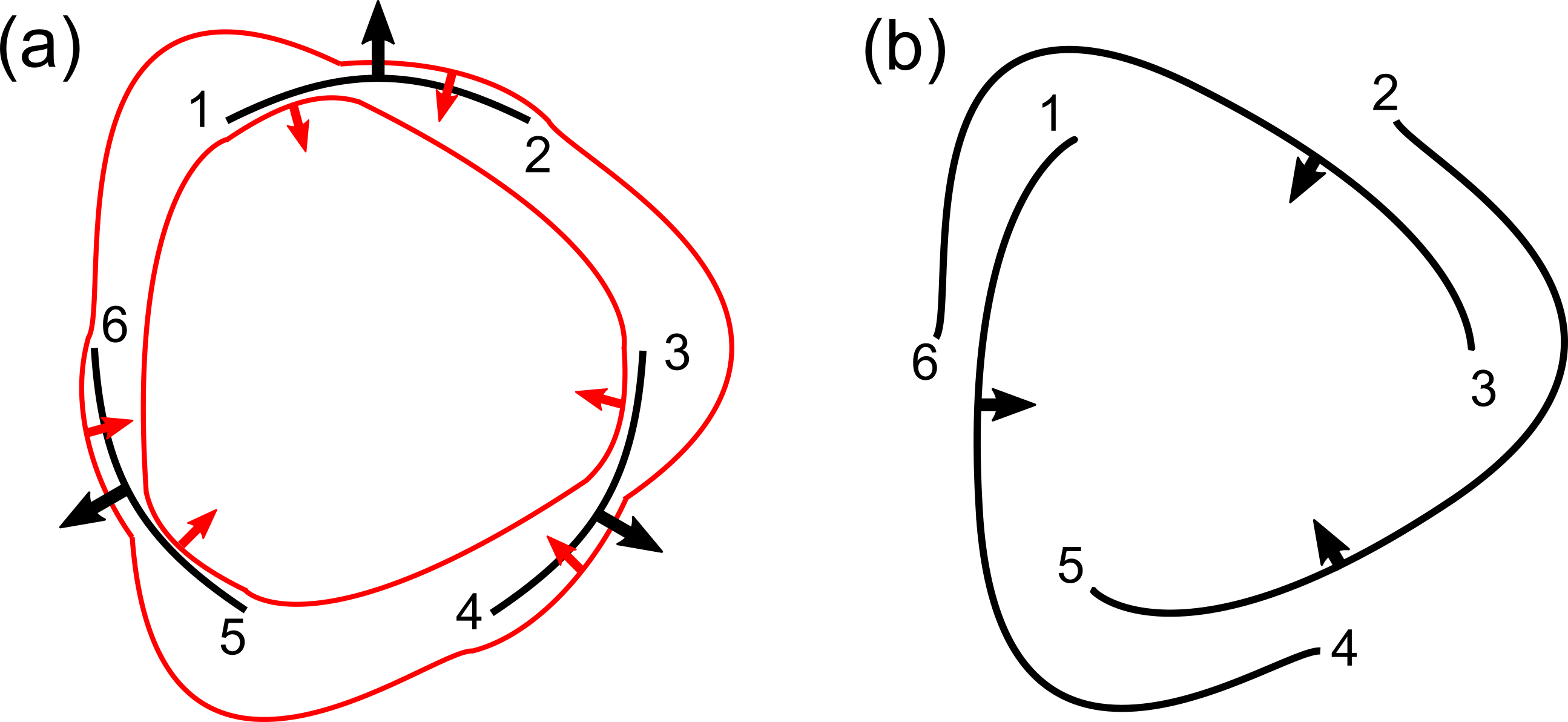}
\caption{(a) Coupling the Fermi arcs on the Weyl semimetal top surface (black) with Fermi pockets of additional 2d systems (red) can lead to a change of Fermi arc connectivity and transform conventional Weyl orbits into nontrivial cyclotron orbit knots in (b). The arrows denote the Fermi velocity directions.}
\label{fig:app4}
\end{figure}

Finally, we discuss the realization of cyclotron orbit knot. One potential scheme is to change the top surface Fermi arc connectivity of an existing Weyl semimetal via proper surface design, such as depositing an adlayer with designated 2D dispersion that couples to the original Fermi arcs, see Fig. \ref{fig:app4}. We note that the reasoning can be generalized to materials with more than three pair of Weyl nodes and surface Fermi arcs, where similar transformation can be adopted for any three chosen Fermi arcs.

\end{document}